\documentclass[12pt]{article}
\pdfoutput=1
\usepackage{tikz}
\usetikzlibrary{matrix,arrows,decorations.pathmorphing,decorations.pathreplacing}
\usepackage{jheppub}
\usepackage{amsmath,amssymb,euscript,array,mathrsfs,appendix,ctable,marvosym}
\usepackage{arydshln}
\usepackage{todonotes}
\usepackage{graphicx}
\usepackage[normalem]{ulem}

\newcommand\blank[1]{#1}
\renewcommand\blank[1]{}
\def\Buildrel#1\over#2\under#3{\mathrel{\mathop{\kern0pt
#2}\limits^{#1}_{#3}}}

\def\CF{{\cal F}}

\def\PP{{\mathbb P}}

\def\mf{{\mathfrak f}}
\def\SO{\text{SO}}

\def\JJ{\mathscr{J}}

\def\mg{\mathfrak g}

\newcommand{\Tr}{\operatorname{Tr}}
\newcommand{\STr}{\operatorname{Str}}

\newcommand{\ad}{\operatorname{ad}}
\newcommand{\Ad}{\operatorname{Ad}}

\def\B0{{\boldsymbol 0}}

\def\SU{\text{SU}}

\def\Dbarslash{\,\,{\raise.15ex\hbox{/}\mkern-12mu {\bar D}}}
\def\Dslash{\,\,{\raise.15ex\hbox{/}\mkern-12mu D}}
\def\delslash{\,\,{\raise.15ex\hbox{/}\mkern-9mu \partial}}
\def\delbarslash{\,\,{\raise.15ex\hbox{/}\mkern-9mu {\bar\partial}}}

\def\LAG{\mathscr{L}}

\newcommand{\MAT}[1]{\begin{pmatrix} #1\end{pmatrix}}

\newcommand{\EQ}[1]{\begin{equation}\begin{split} #1
\end{split}\end{equation}}

\title{Beta Function of k Deformed AdS$_{\boldsymbol 5}\boldsymbol\times \boldsymbol S^{\boldsymbol 5}$ String Theory}
\author{Calan Appadu and Timothy J. Hollowood}
\affiliation{Department of Physics, Swansea University, Swansea, SA2 8PP, U.K.}

\emailAdd{t.hollowood@swansea.ac.uk}

\abstract{We calculate the one loop beta function for the would-be marginal coupling on the world sheet of the $k$ deformed sigma models associated to a quantum group with $q=e^{i\pi/k}$. This includes the bosonic principal chiral models and symmetric space sigma models but also the $k$ deformed semi-symmetric space sigma model describing strings in a deformation of $\text{AdS}_5\times S^5$. The world sheet sigma model is a current-current deformation of the gauged WZW model for the supergroup $\text{PSU}(2,2|4)$ with level $k$. In the string theory context the beta function is shown to vanish because of the vanishing of the Killing form of $\text{PSU}(2,2|4)$ which is another piece of evidence that the $k$ deformed theories define consistent string theories.}

\notoc
\begin{document}

\maketitle

\newpage

\pgfdeclarelayer{top layer}
\pgfdeclarelayer{foreground layer}
\pgfdeclarelayer{background layer}
\pgfsetlayers{background layer,main,foreground layer,top layer}

\section{Introduction}\label{s1}

It is an important question to ask how the gauge-gravity correspondence can be deformed whilst maintaining some of its special features. One such special feature is its integrability (reviewed in \cite{Beisert:2010jr,Arutyunov:2009ga}) and one can ask what kind of deformations of the classic $\text{AdS}_5\times S^5$ background preserve the integrability of the world sheet theory. 

The world sheet theory in the Green-Schwarz formalism for strings on $\text{AdS}_5\times S^5$ written down by Metsaev and Tseytlin \cite{Metsaev:1998it} is a generalized sigma model with a target that is the semi-symmetric space \cite{Serg}  and can be described as the quotient of a Lie supergroup by a bosonic subgroup: 
\EQ{
\frac{\text{PSU}(2,2|4)}{\SO(1,4)\times\SO(5)}\ .
} 
The spectrum of the theory can be found around the vacuum which corresponds to a point-like string orbiting the equator of the $S^5$. Even though the theory is not relativistically invariant, standard integrability arguments allow one to infer the form of its factorizable S-matrix. 
The S-matrix is then the starting point for calculating the energies of string states via Bethe Ansatz techniques. 

The S-matrix of the world sheet theory is associated to a particular rational $R$-matrix solution of the Yang-Baxter equation with an associated Yangian symmetry. This $R$-matrix has a trigonometric generalization associated to a quantum group symmetry with deformation parameter $q$, found by Beisert and Koroteev \cite{Beisert:2008tw}. This trigonometric solution  can be used to define a deformation of the string theory at the level of the world sheet scattering theory \cite{Hoare:2011wr,Hoare:2012fc}. There are two classes of these deformations depending on whether $q$ is real, in which case $q=e^{-\eta}$ (at the classical level) known as the ``$\eta$ deformation", or $q=e^{i\pi/k}$ for an integer $k$, known as the ``$k$ deformation".\footnote{Note that ``$\lambda$ deformation" is not really an appropriate term because the deformation parameter is actually $q$, that is $k$. On the contrary $\lambda$ is the would-be marginal coupling on the world sheet corresponding to the analogue of the 't~Hooft coupling in whatever the dual gauge theory becomes. There is, however, one sense in which $\lambda$ {\it is\/} a deformation parameter. This is in the classical limit $k\to\infty$. In that limit, $\lambda$ acts as the deformation parameter of the Poisson brackets.} 
The two deformation are very different in character. The former, investigated in \cite{Kawaguchi:2012ve,Arutyunov:2013ega,Arutynov:2014ota,Arutyunov:2014cda,Delduc:2014kha,Hoare:2014pna,Vicedo:2015pna,Arutyunov:2015qva} can be interpreted as a direct target space deformation while the latter is more subtle being a discrete deformation. In the latter case, the S-matrix is built from the $R$-matrix in the IRF/RSOS formulation. The $k$-deformation has been investigated in \cite{Arutyunov:2012zt,Arutyunov:2012ai,Hollowood:2014qma,Sfetsos:2014cea,Demulder:2015lva,Hoare:2014pna,Hoare:2015gda,Vicedo:2015pna,Sfetsos:2015nya,Hollowood:2015dpa}.

The big question is whether the $\eta$ or $k$ deformations describe consistent string theories. For the $k$ deformation the evidence is as following:

\begin{enumerate}
\item As described above, the string world sheet theory has a consistent S-matrix \cite{Hoare:2011wr,Hoare:2012fc,Hoare:2013ysa} that describes the scattering of a finite number of states which form the starting point for the TBA \cite{Arutyunov:2012zt,Arutyunov:2012ai}.
\item It is conjectured that the world sheet theory has a consistent Green Schwarz Lagrangian formulation as a deformation of a gauged WZW theory
associated to the supergroup $\text{PSU}(2,2|4)$ \cite{Hollowood:2014qma}. This formulation has the requisite number of kappa symmetries.  
\item The target space geometry satisfies the generalized Einstein equations when a suitable ansatz is made for the RR flux and dilaton based on the bosonic truncation of the Green-Schwarz sigma model mentioned above \cite{Demulder:2015lva}.\footnote{It is not entirely clear whether the solution actually corresponds to the world sheet action. This is discussed \cite{Hoare:2015gda}.}
\end{enumerate}

It is the purpose of the present work to add to this body of evidence by showing that the continuous coupling of the world sheet theory is a marginal coupling in the one loop approximation. This is a world sheet calculation that complements the target space calculation listed above and provides additional evidence in favour of the Green-Schwarz Lagrangian formulation.

This paper is organized as follows. In section \ref{s3}, we lay the ground work for later sections by calculating the beta functions of the integrable sigma models. We use the background field method and choose a particularly simple kind of background field. The novelty of our approach is that we calculate the fluctuations around the background field in terms of the currents of the theories rather than their fundamental field directly. In section \ref{s4} we turn to the $k$ deformed versions of the integrable sigma models. After showing how they have the same equations of motion of the undeformed theories, the beta functions follow in a simple way. Section \ref{s5} then extends the calculation to the semi-symmetric sigma models and their $k$ deformations. We show that the beta functions in both vanish when the group has vanishing Killing form (or, equivalently, dual Coxeter number or quadratic Casimir of the adjoint representation).

\section{Integrable Sigma Models}\label{s3}

In this section, as a necessary precursor to considering the deformed theories, 
we calculate the beta functions of the integrable sigma models. These are the Principal Chiral Models (PCM) and Symmetric Space Sigma Models (SSSM). The PCM can also be formulated as a SSSM associated to Type II symmetric spaces as well.

Our approach is to pick a specific background field. The beta function then follows from computing the spectrum of fluctuations about the background field. Of course this is not a new calculation; however, the novelty of our approach, which will pay dividends later, is to calculate the fluctuations at the level of the currents rather than the fundamental field.

\subsection{Principal chiral models}\label{s3.1}

In this section, we consider the PCM which are sigma models with target spaces equal to a group manifold $F$.
These theories are formulated in terms of a field $f(x,t)$ valued in a Lie group $F$ and an action\footnote{We take 2d metric $\eta_{\mu\nu}=\text{diag}(1,-1)$. We often use the null coordinates $x^\pm=t\pm x$ and for vectors we have $A^\pm=A^0\pm A^1$ and $A_\pm=(A_0\pm A_1)/2$ so that the invariant $A_\mu B^\mu=2(A_+B_-+A_-B_+)$.} 
\EQ{
S[f]=-\frac{\kappa^2}{8\pi t_N}\int d^2x\,\Tr_N\big(f^{-1}\partial_\mu f\ f^{-1}\partial^\mu f\big)\ ,
}
where the trace is taken in the defining representation of dimension $N$ whose Dynkin index is $t_N$: see the appendix for our conventions. Of course, we could use any other faithful representation to define the theory. 

The equations of motion are simply the conservation condition
\EQ{
\partial_+ J_-+\partial_-J_+=0\ ,
\label{em1}
}
for the current  $J_\mu=f^{-1}\partial_\mu f$. This current also satisfies, by virtue of its definition, 
the Cartan-Maurer identity
\EQ{
\partial_+J_--\partial_- J_++[J_+,J_-]=0\ .
\label{cm1}
}
Taken together, \eqref{em1} and \eqref{cm1} can be written as a Lax equation, that is the flatness condition
\EQ{
[\partial_++{\mathscr L}_+(z),\partial_-+{\mathscr L}_-(z)]=0\ ,
\label{leq2}
}
for a $z$-dependent gauge field
\EQ{
{\mathscr L}_\pm(z)=\frac{z}{z\pm1}J_\pm\ ,
}
where $z$, the spectral parameter, is an arbitrary parameter. A Lax representation like this is sufficient to imply the classical integrability of the principal chiral models.

Using \eqref{em1} and \eqref{cm1}, we can write
\EQ{
\partial_\mp J_\pm=\pm\frac12[J_+,J_-]\ .
\label{h45}
}

In order to calculate the beta function, we need to identify a suitable classical background. 
To this end, we take a classical solution of the form
\EQ{
f=\exp\big[x^\mu\Theta_\mu]\ ,
}
where $\Theta_\mu$ are two constant elements of the algebra which commute:
\EQ{
\partial_\mu\Theta_\nu=0\ ,\qquad [\Theta_\mu,\Theta_\nu]=0\ .
}

The Lagrangian evaluated on the background field is simply
\EQ{
\LAG^{(0)}=-\frac{\kappa^2}{8\pi t_N}\Tr_N(\Theta\cdot\Theta)\ .
}

In order to calculate the beta function, we need to extract the operator that governs the 
fluctuations around the background field and calculate its determinant. We will do this at the level of the current $J_\mu$, with its equations of motion \eqref{h45}, rather than the group valued field $f$. Taking the variation of these equations, we have
\EQ{
\partial_\mp\hat J_\pm=\pm\frac12[\Theta_+,\hat J_-]\mp\frac12[\Theta_-,\hat J_+]\ .
}
where we denote a fluctuation by a hat.

So the operator that governs the fluctuations is
\EQ{
{\cal D}=\MAT{\partial_-+\frac12\Theta_- & -\frac12\Theta_+\\ -\frac12\Theta_- &
\partial_++\frac12\Theta_+}
}
acting on $(\hat J_+,\hat J_-)\in(\mf,\mf)$. Note that $\Theta_\pm$ in the above, is given by matrix multiplication in the adjoint representation:\footnote{Note that $\Theta=\Theta^aT_a$ for a set of Hermitian generators $T_a$. Therefore, $[\Theta,T_b]=i\Theta^a f_{ab}{}^cT_c=(\Theta^a)_b{}^cT_c$. See the appendix for our group theory and Lie algebra conventions.}
\EQ{
\big(\Theta_\pm\big)_b{}^c=i\Theta_\pm{}^a\,f_{ab}{}^c\ .
}

After Wick rotation to Euclidean space, the contribution of the fluctuations to the one loop effective Lagrangian is\footnote{In Euclidean space $p_\pm=
(ip_0\pm p_1)/2$.}
\EQ{
\LAG_\text{E}^\text{eff}=
\frac{\kappa^2}{8\pi t_N}\Tr_N(\Theta\cdot\Theta)
+\frac12\int^\mu\frac{d^2p}{(2\pi)^2}\,\Tr\log \MAT{p_-+\frac12\Theta_- & -\frac12\Theta_+\\ -\frac12\Theta_- &
p_++\frac12\Theta_+}\ .
}

The beta function only depends on the 
logarithmically divergent term in the above, which is
\EQ{
-\frac14\int^\mu \frac{d^2p}{(2\pi)^2}\,\frac1{p^2}\Tr_\text{adj}(p\cdot\Theta)^2
=-\frac1{16\pi}\Tr_\text{adj}(\Theta\cdot\Theta)\log\mu+\cdots\ .
}
Note that for consistency the eigenvalues of $\Theta\cdot\Theta$ should be $\leq0$, so the fluctuations have effective masses $\geq0$. 

In the adjoint representation,
\EQ{
\Tr_\text{adj}(\Theta\cdot\Theta)=-\Theta^{\mu\, a}f_{ab}{}^c \,\Theta_\mu{}^d f_{dc}{}^b
=\frac{c_2(F)}{t_N}\Tr_N(\Theta\cdot\Theta)\ ,
\label{nss}
}
where $c_2(F)\equiv t_\text{adj}$ is the quadratic Casimir in the adjoint representation. This is also equal to the dual Coxeter number $h^*$ of $F$.

The RG equation which yields the beta function follows from demanding that the sum of the tree and one loop contributions to the effective action is independent of the cut off scale $\mu$; that is
\EQ{
\mu\frac d{d\mu}\Big[\frac{\kappa^2}{8\pi t_N}\Tr_N(\Theta\cdot\Theta)-\frac{c_2(F)}{16\pi t_N}\Tr_N(\Theta\cdot\Theta)\log\mu\Big]=0\ .
\label{rg1}
}
This yields the beta function of the sigma model coupling 
\EQ{
\mu\frac{d\kappa^2}{d\mu}=\frac{c_2(F)}{2}\ .
\label{be1}
}
Note that, as it should, the beta function does not depend on the choice of background field.

\subsection{Symmetric Space Sigma Models}\label{s3.2}

Symmetric spaces are special quotients of Lie groups $F/G$. One of their defining features is the existence of a $\mathbb Z_2$ automorphism of $\mathfrak f$ under which $\mf=\mf^{(0)}\oplus\mf^{(1)}$, where $\mf^{(0)}\equiv\mg$ is the Lie algebra of the subgroup $G\subset F$. We will denote a decomposition of any element of $a\in\mf$ as $a^{(0)}+a^{(1)}$. The Lie algebra $\mf$ respects the $\mathbb Z_2$ grading:
\EQ{
[\mf^{(i)},\mf^{(j)}]\subset f^{(i+j\ \text{mod}\,2)}\ .
}

Symmetric space are classified as being Type II, of the form $G\times G/G$ (so $F=G\times G$), or Type I, for which $F$ is simple \cite{Hel}. 

We can define sigma models on a symmetric space by a gauging procedure. That is, we write a sigma-model for an $F$-valued field $f(x,t)$ and then gauge the subgroup $G\subset F$ which acts by right-multiplication $f\to fU$, $U\in G$. To this end we introduce a $\mathfrak g$-valued gauge field $B_\mu$ and write
\EQ{
S[f,B_\mu]=-\frac{\kappa^2}{8\pi t_N}\int d^2x\,\text{Tr}_N\big(J^\mu J_\mu\big)\ ,
\label{n3b}
}
where $J_\mu=f^{-1}\partial_\mu f-B_\mu$. The theory is invariant under a global $F$ left action $f\to Uf$, $U\in F$, in addition to the gauge symmetry acting to the right
\EQ{
f\to fU\ ,\qquad B_\mu\to U^{-1}\partial_\mu U+U^{-1}B_\mu U\ .
}

Returning to the sigma model \eqref{n3b}, the equation-of-motion of the gauge field $B_\mu$ imposes the constraint
\EQ{
J_\mu^{(0)}=0\qquad\implies\qquad B_\mu=(f^{-1}\partial_\mu f)^{(0)}\ .
}
The equation-of-motion of the group-valued field $f$ can be decomposed according to $\mf^{(0)}\oplus\mf^{(1)}$ as
\EQ{
&\partial_\pm J^{(1)}_\mp+[B_\pm,J^{(1)}_\mp]=0\ ,\\
&\partial_+B_--\partial_-B_++[B_+,B_-]+[J_+^{(1)},J_-^{(1)}]=0\ .
\label{eom16}
}
We will fix the gauge by imposing the covariant gauge fixing condition
\EQ{
\partial_+B_-+\partial_-B_+=0\ .
}

Classical integrability follows from writing the equations \eqref{eom16} in terms of a Lax pair
\EQ{
[\partial_++{\mathscr L}_+(z),\partial_-+{\mathscr L}_-(z)]=0\ ,
\label{leq3}
}
where 
\EQ{
{\mathscr L}_\pm(z)=B_\pm+z^{\pm1}J_\pm^{(1)}\ .
}

Now we turn to the beta function. 
In this case a suitable background field to take is
\EQ{
f=\exp\big[x^\mu\Theta_\mu]\ ,
\label{bgr}
}
where $\Theta_\mu$ are two constant elements of $\mf^{(1)}$ which commute:
\EQ{
\partial_\mu\Theta_\nu=0\ ,\qquad [\Theta_\mu,\Theta_\nu]=0\ .
}
In particular,
\EQ{
B_\mu=0\ ,\qquad J^{(1)}_\mu=\Theta_\mu\ .
}
Note that due to the $\mathbb Z_2$ grading associated to a symmetric space and the fact that $\Theta_\pm\in\mf^{(1)}$
\EQ{
\ad\Theta_\pm:\quad~~\mf^{(0)}\to\mf^{(1)}\ ,\qquad\ad\Theta_\pm:\quad~~\mf^{(1)}\to\mf^{(0)}\ .
}

The classical Lagrangian evaluated on the background field is
\EQ{
\LAG^{(0)}=-\frac{\kappa^2}{8\pi t_N}\Tr_N(\Theta\cdot\Theta)
}
and the variation of the equations of motion and gauge fixing condition are
\EQ{
&\partial_\pm\hat J^{(1)}_\mp+[\hat B_\pm,\Theta_\mp]=0\ ,\\
&\partial_+\hat B_--\partial_-\hat B_++[\hat J_+^{(1)},\Theta_-]+[\Theta_+,\hat J_-^{(1)}]=0\ ,\\
&\partial_+\hat B_-+\partial_-\hat B_+=0\ .
\label{epp}
}
For the purposes of calculating the beta function, we will not need any more of the paraphernalia of gauge fixing: ghosts, etc.   

The fluctuations are governed by the operator
\EQ{
{\cal D}=\MAT{\partial_-&0&0&-\Theta_+\\
0&\partial_+&-\Theta_-&0\\
-\Theta_-&\Theta_+&-\partial_-&\partial_+\\
0&0&\partial_-&\partial_+}
}
acting on $(\hat J_+^{(1)},\hat J_-^{(1)},\hat B_+,\hat B_-)\in(\mf^{(1)},\mf^{(1)},\mf^{(0)},\mf^{(0)})$.

After Wick rotation, the one-loop effective Lagrangian is 
\EQ{
\LAG_\text{E}^\text{eff}=
\frac{\kappa^2}{8\pi t_N}\Tr_N(\Theta\cdot\Theta)
+\frac12\int^\mu\frac{d^2p}{(2\pi)^2}\,\Tr\log\MAT{p_-&0&0&-\Theta_+\\
0&p_+&-\Theta_-&0\\
-\Theta_-&\Theta_+&-p_-&p_+\\
0&0&p_-&p_+}\ ,
}
from which we extract the logarithmically divergent term
\EQ{
-\frac14\int^\mu \frac{d^2p}{(2\pi)^2}\,\frac1{p^2}\big(\Tr^{(0)}+\Tr^{(1)}\big)(\Theta\cdot\Theta)
&=-\frac1{8\pi}\Tr_\text{adj}(\Theta\cdot\Theta)\log\mu+\cdots\\
&=-\frac{c_2(F)}{8\pi t_N}\Tr_N(\Theta\cdot\Theta)\log\mu+\cdots\ .
\label{ji1}
}
The numerical factor here comes as in \eqref{nss}.

The RG equation is therefore
\EQ{
\mu\frac d{d\mu}\Big[\frac{\kappa^2}{8\pi t_N}\Tr_N(\Theta\cdot\Theta)-\frac{c_2(F)}{8\pi t_N}\Tr_N(\Theta\cdot\Theta)\log\mu\Big]=0
}
and, therefore, the beta function has the form
\EQ{
\mu\frac{d\kappa^2}{d\mu}=c_2(F)\ .
\label{be2}
}

\section{$k$-Deformed Sigma Models}\label{s4}

In this section, we first define and then calculate the beta function of the coupling in the $k$ deformed integrable sigma models. 

The idea is to first re-formulate the original sigma model in a first order form in terms of a Lie algebra-valued field $A_\mu$ and a Lie algebra valued Lagrange multiplier field $\nu$; firstly for the PCM:
\EQ{
S=-\frac{\kappa^2}{2\pi t_N}\int d^2x\,\Tr_N\big[A_+A_-+\nu F_{+-}\big]\ ,
\label{psr}
}
where
\EQ{
F_{+-}=\partial_+A_--\partial_-A_++[A_+,A_-]\ ,
}
is the single non-vanishing component of the curvature of $A_\mu$. 
The vanishing of $F_{+-}$ means that $A_\mu$ is pure gauge and that implicitly there exists a group valued field such that $A_\mu=f^{-1}\partial_\mu f$. So the field $A_\mu$ is identified with the current $J_\mu$ in the original formulation. 

For the SSSM case, the analogous first order form is
\EQ{
S=-\frac{\kappa^2}{2\pi t_N}\int d^2x\,\Tr_N\big[A_+^{(1)}A_-^{(1)}+\nu F_{+-}\big]\ ,
\label{psr}
}

The $k$ deformed sigma models are now obtained by replacing the Lagrange multiplier field with a new field $\CF$
valued in the group $F$ and replacing the term involving the Lagrange multiplier with the gauged WZW action for $\CF$ with gauge field $A_\mu$:\footnote{The deformed theories can also be defined by a dual gauging procedure described in  \cite{Sfetsos:2013wia}.}
\EQ{
-\frac{\kappa^2}{2\pi t_N}\int d^2x\,\Tr_N\big[\nu F_{+-}\big]\longrightarrow S_\text{gWZW}[\CF,A_\mu]\ .
}

The $k$ deformed theory has an action which looks like the action of a gauge WZW model, apart from one term that is deformed:
\EQ{
S[\CF ,A_\mu]&=-\frac k{4\pi t_N}\int d^2x\Tr_N\Big[
\CF ^{-1}\partial_+\CF \,\CF ^{-1}\partial_-\CF +2A_+\partial_-\CF \CF ^{-1}\\ &~~\,~~~~~~~~~~~~~~~~
-2A_-\CF ^{-1}\partial_+\CF -2\CF ^{-1}A_+\CF  A_-+2A_+\Omega A_-\Big]
\\ &~~~~~~~~~+\frac k{24\pi t_N}\int d^3x\,\epsilon^{abc}\Tr_N\,\Big[\CF ^{-1}\partial_a\CF \,
\CF ^{-1}\partial_b\CF \,\CF ^{-1}\partial_c\CF \Big]\ .
\label{gWZW}
}
The deformed term involves the quantity $\Omega$ which acts on the Lie algebra in the following way. For the PCM case, it is simply proportional to the identity:
\EQ{
\text{PCM:}\qquad\Omega=\frac1\lambda\ ,\qquad \lambda=\frac{k}{\kappa^2+k}\ ,
}
while for the SSSM case,
\EQ{
\text{SSM:}\qquad\Omega=\mathbb P^{(0)}+\frac1\lambda\mathbb P^{(1)}\ ,\qquad \lambda=\frac{k}{\kappa^2+k}\ ,
}
where $\mathbb P^{(i)}$ are the projectors onto $\mf^{(i)}$, $i=0,1$. In this case, the deformation vanishes in the component $\mg\equiv\mf^{(0)}$ of the algebra, showing that the theory retains a $G\subset F$ gauge symmetry.

The deformation parameter $k$ has to be an integer for the usual topological reasons associated to the existence of a Wess Zumino term.
This action reduced to the conventional gauged WZW action in the limit that $\Omega\to1$. However, it is important to point out that for generic $\lambda$, the field $A_\mu$, in the $k$-PCM, and the component $A^{(1)}_\mu$, in the $k$-SSSM, is not---strictly-speaking---a gauge field, rather it is to be viewed as an auxiliary field. It is only in the $k$-SSSM case that the theory has a genuine gauge symmetry corresponding to the subgroup $G\subset F$ with gauge field $A^{(0)}_\mu$.

The field $A_\mu$ appears quadratically in the action and it can be integrated out by substituting in the solution of its classical equations of motion
\EQ{
A_+&=\big(\Omega-\Ad\CF^{-1}\big)^{-1}\CF^{-1}\partial_+\CF\ ,\\
A_-&=\big(1-\Ad\CF^{-1}\Omega\big)^{-1}\CF^{-1}\partial_-\CF\ .
\label{icr}
}
This yields an effective sigma model with Wess-Zumino term for the group field $\CF$:
\EQ{
S_\text{eff}=-\frac k{4\pi t_N}\int d^2x\,\Tr_N\Big[
\CF ^{-1}\partial_+\CF\big(1-2\big(1-\Ad\CF^{-1}\Omega\big)^{-1}\big)\CF ^{-1}\partial_-\CF\Big]
+S_\text{WZ}+S_\text{dil}\ .
\label{ger}
}
Here, $S_\text{WZ}$ is the original Wess-Zumino term and, re-instating the world-sheet metric, the dilaton term is
\EQ{
S_\text{dil}=\frac1{4\pi}\int d^2x\,\sqrt{h}\ R^{(2)}\phi\ ,\qquad\phi=-\frac12\Tr\log\big(\Ad\CF-\Omega\big)\ .
}

In the $k$-SSSM case, the theory in the form \eqref{ger} would need to be gauge fixed in some consistent way.

What is remarkable, is that the deformed theories, both $k$-PCM and $k$-SSSM, can be written in way that the classical equations of motion are identical to the original sigma models written in terms of the current $J_\mu$ and, in the SSSM case, the gauge field $B_\mu$.
This will prove key to calculating the beta functions of the deformed theories. At this point we need  to consider the two cases separately.

\subsection{$k$ deformed PCM}\label{s4.1}

The equation-of-motion of the group field $\CF$ can be written either as
\EQ{
\big[\partial_++\CF ^{-1}\partial_+\CF +\CF ^{-1}A_+\CF ,\partial_-+A_-\big]=0\ ,
\label{sd2}
}
or, equivalently, by conjugating with $\CF$, as
\EQ{
\big[\partial_++A_+,\partial_--\partial_-\CF \CF ^{-1}+\CF  A_-\CF ^{-1}\big]=0\ .
\label{sd3}
}
Then, using \eqref{icr} in \eqref{sd2} and \eqref{sd3}, leaves us with the pair of equations
\EQ{
&\lambda \partial_+A_--\partial_-A_++[A_+,A_-]=0\ ,\\
&\partial_+A_--\lambda \partial_-A_++[A_+,A_-]=0\ ,
\label{pls}
}
from which we find (for $\lambda\neq1$)
\EQ{
\partial_\mp A_\pm=\pm\frac1{1+\lambda }[A_+,A_-]\ .
}

These are identical to \eqref{h45}, the equations of the PCM, if we identify
\EQ{
J_\mu=\frac2{1+\lambda}A_\mu\ ,
\label{gww}
}
This fact proves that the $k$-PCM theories are classically integrable but it also suggests a simple way to calculate the beta function of the coupling $\lambda$. The idea is to choose a background field for $\CF$ which is equivalent to the background field that we chose in the PCM model. To this end, let us take a background field
\EQ{
\CF=\exp\big[x^\mu\Lambda_\mu\big]\ .
}
It is important that the background of the fundamental field $\CF$ does not depend on the coupling $\lambda$.

Using the equations \eqref{icr}, with $\Omega=1/\lambda$ in this case, we have
\EQ{
A_\pm=\pm\frac\lambda{1-\lambda}\Lambda_\pm
}
and, hence, via the relation \eqref{gww}
\EQ{
J_\pm=\pm\frac{2\lambda}{1-\lambda^2}\Lambda_\pm\ .
}
This implies that, as far as the fluctuations are concerned, the one loop contribution to the effective Lagrangrian is precisely as in section \ref{s3.1} but with
\EQ{
\Theta_\pm=\pm\frac{2\lambda}{1-\lambda^2}\Lambda_\pm\ .
\label{po1}
}
To be clear, the $\lambda$ dependence is carried by $\Theta_\pm$.

The only difference, compared with the calculation of the beta function in section \ref{s3.1}, lies in the tree level term, which calculated from \eqref{ger} yields
\EQ{
\LAG^{(0)}=-\frac k{16\pi t_N}\frac{1+\lambda}{1-\lambda}\Tr_N(\Lambda\cdot\Lambda)\ .
}

Since the one loop contribution is just as in \eqref{rg1} but with $\Theta_\pm$ replaced with $\Lambda_\pm$ according to \eqref{po1}, the RG equation that results is
\EQ{
\mu\frac d{d\mu}\Big[\frac{k}{16\pi t_N}\frac{1+\lambda}{1-\lambda}\Tr_N(\Lambda\cdot\Lambda)+\frac{c_2(F)}{16\pi t_N}\Big(\frac{2\lambda}{1-\lambda^2}\Big)^2\Tr_N(\Lambda\cdot\Lambda)\log\mu\Big]=0\ .
}
The beta function of the coupling $\lambda$ follows as
\EQ{
\mu\frac{d\lambda}{d\mu}=-\frac{2c_2(F)}k\Big(\frac\lambda{1+\lambda}\Big)^2\ .
}

As a check on this result, we can compare to the analysis of  Tseytlin \cite{Tseytlin:1993hm} who calculated the beta function for a model of this type with completely general symmetric $\Omega$, $\Omega^T=\Omega$, and also 
Sfetsos and Siampos \cite{Sfetsos:2014jfa} who further generalized the analysis to lift the symmetry requirement on $\Omega$. We find complete agreement.

Note that in the $k$ deformed theories, the loop counting parameter is $1/k$, so the result above is exact in $\lambda$ to order $1/k$. In the limit $k\to\infty$, we have
\EQ{
\lambda=1-\frac{\kappa^2}k+\cdots\ ,
}
and the beta function reduces to \eqref{be1}.

\subsection{$k$ deformed SSSM}\label{s4.2}

In the SSSM case, the equations of motion \eqref{sd2}, or equivalently \eqref{sd3}, can, using the constraints \eqref{icr}, be written as 
\EQ{
\partial_\mp A^{(1)}_\pm=[A^{(1)}_\pm,A^{(0)}_\mp]\ ,
\label{yqq}
}
along with 
\EQ{
\partial_+A_-^{(0)}-\partial_-A_+^{(0)}+[A_+^{(0)},A_-^{(0)}]+\lambda^{-1}[A_+^{(1)},A_-^{(1)}]=0\ .
\label{hy2}
}
These equations as precisely equivalent to the equations of the undeformed theory \eqref{eom16} with the identifications 
\EQ{
 B_\pm=A^{(0)}_\pm\ ,\qquad J^{(1)}_\pm=\frac1{\sqrt\lambda}A^{(1)}_\pm\ .
\label{rel2}
}

This fact proves the classical integrability of the deformed theory and, as in the last section, provides a simple way to infer the one loop divergent contribution to the effective Lagrangian. 

For the moment, we shall consider the case of Type II symmetric spaces corresponding to coset $F/G$ with simple $F$.
As before, let us take a background field
\EQ{
\CF=\exp\big[x^\mu\Lambda_\mu\big]\ ,
}
where, more specifically, $\Lambda_\pm\in\mf^{(1)}$. The Lagrangian evaluated on the background field is simply
\EQ{
\LAG^{(0)}=-\frac{k}{16\pi t_N}\frac{1+\lambda}{1-\lambda}\Tr_N(\Lambda\cdot\Lambda)\ .
}

Using the equations \eqref{icr}, with $\Omega=\mathbb P^{(0)}+\lambda^{-1}\mathbb P^{(1)}$, for the background field we have
\EQ{
A_\pm=\pm\frac\lambda{1-\lambda}\Lambda_\pm
}
and hence, via the relation \eqref{rel2},
\EQ{
J_\pm=\pm\frac{\sqrt\lambda}{1-\lambda}\Lambda_\pm\ ,\qquad B_\pm=0\ .
}
Comparing with our analysis in section \ref{s3.2}, implies that the one 
loop contribution to the effective Lagrangrian is precisely as derived there but with 
\EQ{
\Theta_\pm=\pm\frac{\sqrt\lambda}{1-\lambda}\Lambda_\pm\ .
\label{po2}
}
Again, the $\lambda$ dependence is carried by $\Theta_\pm$. In addition, it is important that the fluctuations are stable around the background which requires that
$\Theta\cdot\Theta$ is a negative operator, i.e.~$\Lambda\cdot\Lambda$ is positive operator. 

Hence, the RG equation in the deformed theory takes the form
\EQ{
\mu\frac d{d\mu}\Big[\frac{k}{16\pi t_N}\frac{1+\lambda}{1-\lambda}\Tr_N(\Lambda\cdot\Lambda)+\frac{c_2(F)}{8\pi t_N}\Big(\frac{\sqrt\lambda}{1-\lambda}\Big)^2\Tr_N(\Lambda\cdot\Lambda)\log\mu\Big]=0\ ,
}
from which we extract the beta function
\EQ{
\mu\frac{d\lambda}{d\mu}=-\frac{c_2(F)}k\lambda\ .
}
In the limit $k\to\infty$, we have
\EQ{
\lambda=1-\frac{\kappa^2}k+\cdots
}
and the beta function reduces to \eqref{be2}.

The result here can also be extracted from the analysis of Tseytlin \cite{Tseytlin:1993hm} and Sfetsos and Siampos \cite{Sfetsos:2014jfa} (see also \cite{Itsios:2014lca}) by taking the result of the un-gauged theory with
\EQ{
\Omega=\frac1{\lambda_0}\mathbb P^{(0)}+\frac1\lambda\mathbb P^{(1)}
}
and then taking the limit $\lambda_0\to1$. Note that in the un-gauged theory $\lambda_0$ also runs and, as noted in \cite{Sfetsos:2014jfa}, $\lambda_0=1$ is not fixed point, unless $\mg$ is abelian. However, one should remember that a gauge symmetry implies a redundancy in the description and unless there is a gauge anomaly quantum effects cannot break the symmetry. Hence, $\lambda_0$ is not a running coupling. Ultimately, however, it is not entirely obvious that the result in a gauge theory can be extracted from a theory without a gauge symmetry by taking a limit. Nevertheless this appears to be true in the present context.\footnote{We suspect that one gets the correct result at the one loop level because the degrees of freedom that should be gauged away do not couple to the background field at this order because of the special properties of the structure constants of $F$ in a basis aligned with $G\subset F$ and the the fact that the latter is a symmetric space.}

For the Type II cases there is a generalization. Because $F=G\times G$ is not simple, the deformed version of the sigma model can have separate levels $k_1$ and $k_2$ for each group factor. The only effect of this generalization is to change the tree level Lagranagian of the background field to 
\EQ{
\LAG^{(0)}=-\frac{k_1+k_2}{16\pi t_N}\frac{1+\lambda}{1-\lambda}\Tr_N(\Lambda\cdot\Lambda)
}
and to replace the quadratic Casimir $c_2(F)\to c_2(G)$ leading to a beta function
\EQ{
\mu\frac{d\lambda}{d\mu}=-\frac{c_2(G)}{k_1+k_2}\lambda\ .
\label{gtt}
}

We can also compare the result \eqref{gtt} with a CFT analysis. This is pertinent to the limit $\lambda\to0$ where, by integrating out the components $A^{(1)}_\mu$, the deformed theory can be interpreted as a current-current deformation of the gauged $F/G$ WZW model
\EQ{
S=S_\text{gWZW}[\CF,A_\mu^{(0)}]+\frac{4\pi \lambda}{k}\int d^2x\,\text{Tr}\,\big(\hat\JJ_+^{(1)}\hat\JJ_-^{(1)}\big)+\cdots\ .
}
In the above, $\hat\JJ_\pm$ are the usual Kac-Moody currents of the $F/G$ gauged WZW model
\EQ{
\hat\JJ_+&=-\frac k{2\pi}\big(\CF^{-1}\partial_+\CF+\CF^{-1}A_+^{(0)}\CF-A_-^{(0)}\big)\ ,\\
\hat\JJ_-&=\frac k{2\pi}\big(\partial_-\CF\CF^{-1}-\CF A_-^{(0)}\CF^{-1}+A_+^{(0)}\big)\ .
}
The deformed CFT has been analysed in \cite{Ahn:1990gn}. The deforming operator, involving, as it does, a product of currents in the coset directions, na\"\i vely looks like it should be marginal. However, as an operator in the CFT it must be rendered properly gauge invariant by dressing with Wilson lines of the gauge field. This modifies the  na\"\i ve conformal dimension from $(1,1)$ to $(\Delta,\Delta)$ with
\EQ{
\Delta=1-\frac{h^*}{k_1+k_2+h^*}\ ,
}
where $h^*$ is the dual Coxter number. So the deformation is actually relevant. The anomalous dimension in the limit $k_1,k_2\gg h^*$ matches \eqref{gtt} exactly, using the fact that $c_2(F)\equiv h^*$.

\section{The Semi-Symmetric Space Sigma Models}\label{s5}

We now consider the generalization to a semi-symmetric space. 
A semi-symmetric space $F/G$ is a generalization of a concept of a symmetric space to a supergroup. So $F$ is a supergroup, like the quotient $G\subset F$ is a particular bosonic subgroup. If $F_B$ is the bosonic subgroup of $F$ then $F_B/G$ is an ordinary symmetric space. What defines a semi-symmetric space is the fact that the super Lie algebra $\mf$ admits at $\mathbb Z_4$ grading $\mf=\oplus_{i=0}^3\mf^{(i)}$ such that
\EQ{
[\mf^{(i)},\mf^{(j)}]\subset\mf^{(i+j\ \text{mod}\ 4)}\ ,
\label{lgr}
}
which generalizes the $\mathbb Z_2$ grading of the ordinary symmetric space. In particular, $\mf^{(0)}=\mg$ and 
$\mf^{(1)}$ and $\mf^{(3)}$ are the Grassmann odd components of the algebra.\footnote{The spaces $\mf^{(i)}$ are defined as linear combinations of generators with coefficients that are Grassmann even or odd depending on whether the generator is even or odd graded.}

The bosonic/fermionic parts of the superalgebra are precisely the even/odd graded components, and $\mf^{(0)}\equiv\mg$ is the Lie algebra of the ordinary Lie group $G$. The supertrace in the defining representation defines a bilinear form on the generators $T^a$:
\EQ{
\STr(T_aT_b)=\eta_{ab}\ .
\label{ip8}
}
We will always take a basis of generators which 
respects the $\mathbb Z_4$ grading and so $\eta_{ab}$ pairs generators of $\mf^{(1)}$ with $\mf^{(3)}$.

The particular semi-symmetric space
\EQ{
\frac FG=\frac{\text{PSU}(2,2|4)}{\SO(1,4)\times\SO(5)}\ ,
}
describes the string world sheet in the Green Schwarz formalism for the $\text{AdS}_5\times S^5$ background. 

The action of the theory has the form
\EQ{
S=-\frac{\kappa^2}\pi\int d^2x\,\STr\Big[J_+^{(0)}J_-^{(0)}+J_+^{(2)}J^{(2)}_--\frac12J^{(1)}_+J^{(3)}_-+\frac12J^{(1)}_-J^{(3)}_+\Big]\ ,
\label{pss}
}
where $J_\mu=f^{-1}\partial_\mu f-B_\mu$ where $f$ is a field valued in $F$ and $B_\mu$ is a gauge field valued in the Lie algebra $\mg$. 

The equations-of-motion of the sigma model, along with the Cartan-Maurer identity 
\EQ{
\partial_+J_--\partial_-J_++[J_+,J_-]=0\ ,
}
can be decomposed with respect to the $\mathbb Z_4$ grade as the group of equations
\EQ{
&D_+ J^{(2)}_- +[J_+^{(1)},J_-^{(1)}]=0\ ,\\
&D_-J^{(2)}_+ + [J_-^{(3)},J_+^{(3)}]=0\ ,\\
&\partial_+B_--\partial_-B_++[B_+,B_-]+[J_+^{(2)},
J_-^{(2)}]\\ &\hspace{2.8cm}+[J_+^{(3)},J_-^{(1)}]+[J_+^{(1)},J_-^{(3)}]=0\ ,\\
&D_+J_-^{(1)}-D_-J_+^{(1)}+[J_+^{(3)},J_-^{(2)}]=0\ ,\\
&D_+J_-^{(3)}-D_-J_+^{(3)}+[J_+^{(2)},J_-^{(1)}]=0\ ,\\
&[J_+^{(1)},J_-^{(2)}]=[J_+^{(2)},J_-^{(3)}]=0\ .
\label{eom4}
}
In the above, there is a $\mg$-valued connection $D_\pm\cdot  =[\partial_\pm+B_\pm,\cdot]$. Note that in formulation we have chosen, the components $J_\mu^{(0)}$ valued in $\mg$ vanish by virtue of the equation of motion of the gauge field.

The plethora of equations \eqref{eom4} can be written compactly in Lax form, which demonstrates integrability at the classical level \cite{Bena:2003wd,Alday:2005gi}:
\EQ{
[\partial_\mu+{\mathscr L}_\mu(z),\partial_\nu+{\mathscr L}_\nu(z)]=0\ ,
\label{leq2}
}
with
\EQ{
{\mathscr L}_\pm(z)=B_\pm+z J_\pm^{(1)}+z^{\mp2} J_\pm^{(2)}+z^{-1} J_\pm^{(3)}\ ,
\label{rmm}
}
where $z$, the spectral parameter, is an arbitrary parameter.

\subsection{The $k$ Deformation}

We can attempt to define a $k$ deformed version of the theory in the same way as in the bosoonic theories in section \eqref{s3}. To this end, one writes the original sigma model in first order form
\EQ{
S=-\frac{\kappa^2}{2\pi t_N}\int d^2x\,\STr_N\Big[A_+^{(2)}A_-^{(2)}-\frac12A^{(1)}_+A^{(3)}_-+\frac12A^{(1)}_-A^{(3)}_++\nu F_{+-}\Big]\ .
\label{psr2}
}
However, simply replacing the final term by the gauged WZW action for a field $\CF$ valued in the supergroup $F$ does not work in this case in the sense that integrability is lost. The same fate befalls Sfetsos's dual gauging procedure  \cite{Sfetsos:2013wia}.

In order to preserve integrability, the coefficients of the first three terms in \eqref{psr2} must be re-weighted in a very particular way \cite{Hollowood:2014qma}, to wit, 
\EQ{
&\kappa^2\Big[A_+^{(2)}A_-^{(2)}-\frac12A^{(1)}_+A^{(3)}_-+\frac12A^{(1)}_-A^{(3)}_+\Big]\\
&\longrightarrow k\Big[\Big(\frac1{\lambda^2}-1\Big)A_+^{(2)}A_-^{(2)}+\Big(\lambda-1\Big)A^{(1)}_+A^{(3)}_-+\Big(\frac1{\lambda}-1\Big)A^{(1)}_-A^{(3)}_+\Big]\ ,
}
for some parameter $\lambda$.
The re-weighting should go away in the limit $k\to\infty$ which implies
\EQ{
\lambda=1-\frac{\kappa^2}{2k}+{\cal O}(1/k^2)\ .
}

After the re-weighting, the final term in \eqref{psr2} is replaced with the gauged WZW action to give the deformed theory:
\EQ{
S[\CF ,A_\mu]&=-\frac k{2\pi}\int d^2x\STr\Big[
\CF ^{-1}\partial_+\CF \,\CF ^{-1}\partial_-\CF +2A_+\partial_-\CF \CF ^{-1}\\ &~~~~~~~~~~~~~
-2A_-\CF ^{-1}\partial_+\CF -2\CF ^{-1}A_+\CF  A_-+2A_+\Omega A_-\Big]
\\ &~~~~~~~~+\frac k{12\pi}\int d^3x\,\epsilon^{abc}\STr\,\Big[\CF ^{-1}\partial_a\CF \,
\CF ^{-1}\partial_b\CF \,\CF ^{-1}\partial_c\CF \Big]\ ,
\label{gWZW2}
}
where
\EQ{
\Omega&=\PP^{(0)}+\frac{1}\lambda\PP^{(1)}+\frac{1}{\lambda^2}\PP^{(2)}+\lambda\PP^{(3)}\ ,\\
\Omega^T&=\PP^{(0)}+\lambda\PP^{(1)}+\frac{1}{\lambda^2}\PP^{(2)}+\frac{1}\lambda\PP^{(3)}\ ,
}
so that $\STr(A_+\Omega A_-)=\STr(A_-\Omega^T A_+)$. WZW theories on coset superspaces have been studied in \cite{Quella:2013oda}.

The form of $\Omega$ is completely fixed by requiring that the deformed theory is also integrable.
This follows because, in addition to be constraints \eqref{icr}, the equations of motion of theory
are identical to the original sigma model with the relations
\EQ{
A_\pm^{(0)}&=B_\pm\ ,\qquad~\,  A_\pm^{(1)}=\lambda^{\mp1/2}J_\pm^{(1)}\ ,\\
A_\pm^{(2)}&=\lambda J_\pm^{(2)}\ ,\qquad A_\pm^{(3)}=\lambda^{\pm1/2}J_\pm^{(3)}\ .
\label{rel}
}

It is worth point out, that, contrary to the bosonic case, the limit $\lambda\to0$ does not describe a small deformation of the gauged $F/G$ WZW by a current-current operator; in fact, as $\lambda\to0$,
\EQ{
S=S_\text{gWZW}[\CF,A_\mu^{(0)}]+
\frac{4\pi}{k}\int d^2x\,\text{Tr}\,\big(-\hat\JJ_+^{(1)}\hat\JJ_-^{(3)}+\lambda^2\hat\JJ_+^{(2)}\hat\JJ_-^{(2)}+\lambda\hat\JJ_+^{(3)}\hat\JJ_-^{(1)}\big)+\cdots\ .
}
So the UV limit is not just a gauged WZW model. This latter point is important because it implies that the fermionic kinetic terms are like those of a Green-Schwarz sigma model rather than a WZW model.

\subsection{Beta function}

Before proceeding, let us first consider the sigma model from the point of view of a string world theory for which there are 
some complications. In particular, one must introduce a world sheet metric. Choosing conformal gauge, the metric effectively disappears but leaves its equation of motion in the form of the Virasoro constraints. The theory also has a local fermionic symmetry, kappa symmetry, whose effect is to reduce the number of degrees of freedom in the fermionic sector so as to ensure spacetime supersymmetry. 

However, although, these details are crucial they do not really affect the one-loop beta function calculation to which we now turn. Background field calculations of the beta functions in these kinds of theory have been done by Polyakov  \cite{Polyakov:2004br}, Babichenko \cite{Babichenko:2006uc}, Adam et al \cite{Adam:2007ws}
 and Zarembo \cite{Zarembo:2010sg}. 
   
Following the previous logic, we now settle on a suitable background field. The idea is to choose 
a bosonic configuration as in \eqref{bgr} with $\Theta_\mu\in\mf^{(2)}$,
\EQ{
f=\exp\big[x^\mu\Theta_\mu]\ .
}
In the deformed theory, one chooses
\EQ{
\CF=\exp\big[x^\mu\Lambda_\mu\big]\ ,
}
which gives
\EQ{
\Theta_\pm=\pm\frac\lambda{1-\lambda^2}\Lambda_\pm\ .
\label{tyy}
}
As previously, both $\Theta_\mu$ and $\Lambda_\mu$ are constant with $[\Theta_+,\Theta_-]=[\Lambda_+,\Lambda_-]=0$.

Note that the bosonic sector of the theory is a product of two symmetric spaces $\text{AdS}_5=\SO(2,4)/\SO(1,4)$ and $S^5=\SO(6)/\SO(5)$ and so we can use the formulae of section \ref{s4.2} but with $\lambda\to\lambda^2$. So, for example, \eqref{tyy} is just \eqref{po2} with $\lambda\to\lambda^2$.
As for the bosonic sigma models, we can treat both the original sigma model and its $k$ deformation at the same time since the equations of motion are identical. The only difference lies in the $\lambda$ dependence of $\Theta_\pm$ in the deformed theory.

The advantage of choosing a purely bosonic background is that the bosonic and fermionic sectors are completely decoupled at one loop order. 
The contribution to the one loop logarithmic divergence from the bosonic fields is identical to that of a $F_B/G$ symmetric space which, before momentum integration was written in \eqref{ji1}. In the present situation, we identify $\Tr_\text{adj}$ of $F_B$ as $\Tr^{(0)}+\Tr^{(2)}$ of the semi-symmetric space. The bosonic contribution is, therefore,
\EQ{
-\frac14\int^\mu \frac{d^2p}{(2\pi)^2}\,\frac1{p^2}\Big[\Tr^{(0)}(\Theta\cdot\Theta)+
\Tr^{(2)}(\Theta\cdot\Theta)\Big]\ .
}

Now we turn to the fermionic sector. The variation of the fermionic equations in \eqref{eom4} gives
\EQ{
&\partial_+\hat J_-^{(1)}-\partial_-\hat J_+^{(1)}+[J_+^{(3)},\Theta_-]=0\ ,\\
&\partial_+\hat J_-^{(3)}-\partial_-\hat J_+^{(3)}+[\Theta_+,J_-^{(1)}]=0\ ,\\
&[\hat J_+^{(1)},\Theta_-]=[\Theta_+,\hat J_-^{(3)}]=0\ .
}
At this point, one might think that one would have to fix the kappa symmetry. However, as remarked by Zarembo \cite{Zarembo:2010sg}, the unphysical modes simply do not couple to the background field and so they do not need to be actually projected out for the purposes of this computation.

The fluctuation operator for the fermionic modes is 
\EQ{
{\cal D}=\MAT{-\partial_-&\partial_+&-\Theta_-&0\\ 0&\Theta_+&-\partial_-&\partial_+\\ -\Theta_-&0&0&0\\ 0&0&0&\Theta_+}
}
acting on $(\hat J^{(1)}_+,\hat J_-^{(1)},\hat J_+^{(3)},\hat J_-^{(3)})$.
Hence, the contribution from the fermionic sector that is relevant to the background field dependence of the logarithmic divergence is  clearly
\EQ{
\frac14\int^\mu \frac{d^2p}{(2\pi)^2}\,\Tr\log\MAT{p_+&-\Theta_-\\ \Theta_+&-p_-}
=\frac1{2\pi}\Big[\Tr^{(1)}(\Theta_-\Theta_+)+
\Tr^{(3)}(\Theta_+\Theta_-)\Big]\log\mu+\cdots\ .
}
Note that modes in $\mf^{(1)}$, respectively $\mf^{(3)}$, that lie in the kernel of $\ad\Theta_+$, respectively $\ad\Theta_-$, do not contribute. These are precisely the unphysical modes that are removed by fixing kappa symmetry.

Hence, adding the bosonic and fermionic contributions together gives the logarithmically divergent term
\EQ{
&-\frac1{8\pi}\Big[\Tr^{(0)}(\Theta\cdot\Theta)-
\Tr^{(1)}(\Theta\cdot\Theta)+\Tr^{(2)}(\Theta\cdot\Theta)-
\Tr^{(3)}(\Theta\cdot\Theta)\Big]\log\mu\\ &=-\frac1{8\pi}\STr_\text{adj}(\Theta\cdot\Theta)\log\mu\ .
}
The factor\footnote{In the following, $|a|=0$ for $T_a\in\mf^{(0)},\mf^{(2)}$ and $|a|=1$ for $T_a\in\mf^{(1)},\mf^{(3)}$.}
 \EQ{
\STr_\text{adj}(\Theta\cdot\Theta)=-\Theta^{\mu\, a}\Theta_\mu{}^d (-1)^{|b|}f_{ab}{}^c\,f_{dc}{}^b\ ,
}
involves the Killing form of $\mf$.

Just as with symmetric space sigma models, the divergent term is proportional to the quadratic Casimir in the adjoint $c_2(F)$, equivalently the dual Coxeter number $h^*$ or the normalization of the Killing form. But in the case of $F=\text{PSU}(2,2|4)$, this vanishes and so the coupling $\lambda$ is marginal to this order. This is obviously a necessary condition that the deformed theory defines a consistent string background.

\section{Discussion}

We have shown how to calculate the one loop beta function of a series of integrable sigma models and their so-called $k$ deformations by using a background field method. The novelty of the approach was to consider the fluctuations around the background field in terms of the currents of the first order formalism. This allowed us to extend our calculations to the $k$ deformed theories in a simple way. 

The key result is that the beta function of the $k$ deformed semi-symmetric space vanishes when it vanishes in the un-deformed theory. This is further evidence that the $k$ deformed semi-symmetric space theory for $\text{PSU}(2,2|4)/\SO(1,4)\times\SO(5)$ is a consistent Green-Schwarz sigma model for superstring theory on a deformation of $\text{AdS}_5\times S^5$. 

The deformed string theory is described by two parameters $(g,k)$, where $k\in\mathbb Z>0$. Here, $g$ is the coupling of the sigma model that for the $\text{AdS}_5\times S^5$ background is the 't~Hooft coupling of the dual gauge theory. The excitations on the world sheet have a non-relativistic dispersion relation that can be written as \cite{Hoare:2011wr,Hoare:2012fc,Hoare:2013ysa,Hollowood:2014rla}
\EQ{
\sin^2\Big(\frac{\xi E}{4g}\Big)-\xi^2\sin^2\Big(\frac p{4g}\Big)=(1-\xi^2)\sin^2\Big(\frac{\pi\EuScript Q}{2k}\Big)\ ,
\label{disp}
}
where, 
\EQ{
\xi=\frac{2g\sin(\pi/k)}{\sqrt{1+4g^2\sin^2(\pi/k)}}
}
and $\EuScript Q=1,2,\ldots,k$ is a quantum number that labels a multiplet of states in the spectrum. The coupling $g$ is related to $\lambda$ used in this work via
\EQ{
\Big(4g\sin\frac\pi k\Big)^2=\frac{(1-\lambda)^2}{4\lambda}\ .
}
The classical limit corresponds to $k\to\infty$, $g\to\infty$ with the ratio fixed. We see then that $\lambda$ parametrizes this ratio. The fact that $\lambda$ is a marginal coupling (to one loop at least) means that $g$ is marginal. 

\vspace{0.2cm}
\begin{center}
{\tiny******}
\end{center}
\vspace{0.2cm}

\noindent TJH would like to thanks Luis Miramontes, David Schmidtt and Ben Hoare for valuable 
discussions and collaboration. We would also like to thank Kostas Siampos and Kostas Sfetsos for pointing out an important typo in an earlier draft of the paper. TJH is supported in part by the STFC grant ST/G000506/1. 
CA is supported by an STFC studentship.

\appendix
\appendixpage

\section{Conventions}

We define the Lie algebra by a set of Hermitian generators
\EQ{
[T_a,T_b]=if_{ab}{}^cT_c\ .
}
If we have a representation of a Lie algebra ${\cal R}$ of a Lie group $F$, we can define the Dynkin index $T_{\cal R}$ and the quadratic Casimir
$c_2(\cal R)$:
\EQ{
\Tr_{\cal R}\big[T_aT_b\big]=T_{\cal R}\eta_{ab}\ ,\qquad \sum_aT_aT_a=c_2({\cal R}){\bf 1}\ ,
}
where for a compact Lie algebra $\eta_{ab}=\delta_{ab}$. It follows that they are related via
\EQ{
T_{\cal R}\,\text{dim}(F)=c_2({\cal R})\,\text{dim}({\cal R})\ .
}
For the adjoint representation, $\text{dim}({\cal R})=\text{dim}(F)$ and
\EQ{
c_2(\text{adj})\equiv c_2(F)=T_\text{adj}\ .
}
In addition, $c_2(F)=h^*$, the dual Coxeter number. This number also defines the 
overall normalization of the Killing form:
\EQ{
\Tr_\text{adj}\big(T_aT_b)=T_\text{adj}\eta_{ab}\ .
}

For $\SU(N)$, $T_\text{adj}=c_2(\SU(N)))=h^*=N$ and $T_N=\frac12$, while for 
$\SO(N)$ $T_\text{adj}=c_2(\SO(N))=h^*=N-2$ and $T_N=1$.

\end{document}